# Generations of high efficiency, high purity, and broadband Laguerre-Gaussian modes from a Janus optical parametric oscillator

Dunzhao Wei[1,†], Pengcheng Chen[1,†], Xiaopeng Hu[1,2,†], Yipeng Zhang[1,†], Wenzhe Yao[1], Rui Ni[1], Xinjie Lv[1,2], Yong Zhang[1,2,*], Shining Zhu[1,2,*], and Min Xiao[1,2,3,*]

[1]National Laboratory of Solid State Microstructures, College of Engineering and Applied Sciences, and School of Physics, Nanjing University, Nanjing 210093, China

[2]Collaborative Innovation Center of Advanced Microstructures, Nanjing University, Nanjing 210093, China

[3]Department of Physics, University of Arkansas, Fayetteville, Arkansas 72701, USA

[†]These authors contribute equally to this work.

*To whom correspondence should be addressed: zhangyong@nju.edu.cn; zhusn@nju.edu.cn; mxiao@uark.edu.

**Abstract**

Laguerre-Gaussian (LG) modes, carrying orbital angular momentum of light, are critical for important applications such as high-capacity optical communications, super-resolution imaging, and multi-dimensional quantum entanglement. Advanced developments in these applications strongly demand reliable and tunable LG mode laser sources, which, however, do not yet exist. Here, we experimentally demonstrate highly-efficient, highly-pure, broadly-tunable, and topological-charge-controllable LG modes from a Janus optical parametric oscillator (OPO). Janus OPO featuring two-face cavity mode is designed to guarantee an efficient evolution from a Gaussian-shaped fundamental pumping mode to a desired LG parametric mode. The output LG mode has a tunable wavelength between 1.5 μm and 1.6 μm with a conversion efficiency above 15%, a topological charge switchable from -4 to 4, and a mode purity as high as 97%, which provides a high-performance solid-state light source for high-end demands in multi-dimensional multiplexing/demultiplexing, control of spin-orbital coupling between light and atoms, and so on.

**Introduction**

Laguerre-Gaussian (LG) modes with unique spiral wavefronts are the paraxial solutions of scalar Helmholtz equation in cylindrical coordinates [1]. In 1992, Allen et al. demonstrated that an LG mode carries an orbital angular momentum (OAM) of $l\hbar$ per photon [2], where $l$ is called the topological charge (TC). Their pioneering work has significantly boosted the applications of LG modes from optical trapping and optical tweezer to optical communications, super-resolution imaging, precision measurement, quantum information processing, and so on [3-10]. In turn, these high-end demands have triggered the developments of LG mode laser sources in recent years [11-15]. Almost all the applications benefit from the high purity of an LG laser source, such as improved signal-to-noise ratio in rotation measurement, enhanced resolution in fluorescence imaging, and optimized coupling with an OAM photonic chip[16-18]. High-power laser output of LG mode could provide an effective way to decrease thermal noises in gravitational-wave detection [19,20]. In particular, LG laser sources are hoped to be wavelength-tunable for wavelength division multiplexing in OAM-based high-capacity optical communication, investigation of spin-orbital coupling with various atoms in quantum storage and isolation, and excitations of versatile fluorescence in super-resolution imaging [17,21-25]. However, a reliable and broadband-tunable LG mode laser source does not exist yet.

Optical parametric oscillator (OPO) has been recognized as one of the most popular tunable sources[26-31]. A pump wave with frequency of $\omega_p$ generates two parametric waves[32], i.e., signal and idler waves at the frequencies of $\omega_s$ and $\omega_i$, respectively, through the second-order nonlinear down-conversion process. It satisfies the energy conservation of $\omega_p = \omega_i + \omega_s$. By controlling the phase matching condition for momentum conservation, one can obtain wavelength-tunable output of the generated parametric waves[33]. An OPO system is capable to output broad wavelengths covering UV, visible, and infrared bands, which provides an excellent candidate for broadband output of high-quality LG modes. There are two reported configurations before. One is to build a traditional OPO outputting a Gaussian mode, and then convert it to an LG mode by using a spiral phase plate, a fork grating, a Q-plate, a vector vortex waveplate (VVW) or a spatial light modulator (Fig. 1a)[7,34-38]. Because these devices only introduce a spatial phase modulation, the generated beam is actually a superposition of various higher-order LG modes with the same azimuthal index $l$ but different radial index $p$, i.e., $\sum_p LG(l,p)$. It suffers from poor mode purity, and generally, the higher TC is, the lower mode purity becomes (typically 80% and 60% for LG(1,0) and LG(2,0) modes, respectively) [39,40]. The other approach is to oscillate an LG mode inside the OPO cavity (Fig. 1b) by utilizing the fact that LG modes are Eigen cavity modes[41,42]. In comparison to a Gaussian-mode OPO system, the frequency conversion of LG mode is less efficient because its donut-shaped profile has a much lower power density. In addition, the output mode quality is not as good as hoped.

Here, we propose and experimentally demonstrate a Janus OPO system for highly-efficient output of highly-pure, broadly-tunable, and TC-controllable LG modes (Fig. 1c). The Janus OPO features the two-face cavity mode, which combines the advantages of both Gaussian and LG cavity modes. The front face at the input mirror has a Gaussian profile to achieve a better conversion efficiency because of its higher power density relative to the LG mode. The back face at the end mirror is a donut-shaped LG profile, which guarantees the direct output of a high-purity LG mode. The key question is how to smoothly evolve the cavity mode from a Gaussian profile to an LG profile, and vice versa, without breaking the cavity mode reversibility. The general idea is to directly put a spatial phase modulator such as a VVW into the cavity to complete the mode conversion[13,43,44]. However, only phase modulation is not sufficient to perform a perfect spatial mode conversion. Let us consider an ideal LG mode at the output mirror. As shown in Fig. 1d, it propagates through the VVW, which produces a beam superimposed by multiple modes in Part I of the Janus OPO rather than a single mode as in a traditional cavity[45]. This superimposed beam of multiple spatial modes hardly keeps its profile during free propagation. Therefore, the VVW alone cannot convert it back into the same LG mode as the initial one (Fig. 1d), which breaks the spatial mode reversibility inside the cavity. Under this situation, low efficiency and poor mode purity are inevitable at the output.

To realize an ideal Janus OPO (Fig. 1c), the mode reversibility has to be simultaneously satisfied for multiple modes in Part I of the cavity. We introduce an imaging system into the cavity. In our experiment, we use a concave front (input) mirror as an equivalent imaging lens for the compact Janus OPO design (Fig. 1c). Figure 1e shows the transformation of Janus cavity mode in a round trip. When the imaging system works properly, the multiple spatial modes well repeat themselves after passing through the equivalent lens (i.e., being reflected back at the concave front mirror). Then, the VVW can convert them back into an ideal LG mode in Part II of the Janus OPO and the reversibility condition inside cavity can therefore be perfectly fulfilled in principle. In addition, the cavity mode profile near the front mirror is required to match the pump Gaussian mode.

In our experiment, the multiple modes after an LG mode passing through the VVW compose a so-called hollow-Gaussian beam[45], which naturally evolves into a spatial profile very close to a Gaussian mode after a certain propagation distance (Fig. 1d). It should be noted that the hollow-Gaussian beam cannot recover itself without using an imaging system (Fig. 1d).

In this work, we experimentally demonstrate a Janus OPO based on quasi-phase-matching (QPM) configuration[26,46-49]. The nonlinear crystal, i.e. a periodically-poled lithium niobate (PPLN) crystal, is set next to the front mirror to fully utilize the Gaussian-like front face of Janus cavity mode for high conversion efficiency. The generated parametric light is naturally converted into a designed LG mode at the output port of the cavity. The experimental results present a high-performance LG mode source beyond the existing methods. For the generated signal LG beam, the wavelength is tunable between 1.5 μm and 1.6 μm, the conversion efficiency is above 15%, the TC is switchable from -4 to 4, and most importantly, the mode purity can reach 97%.

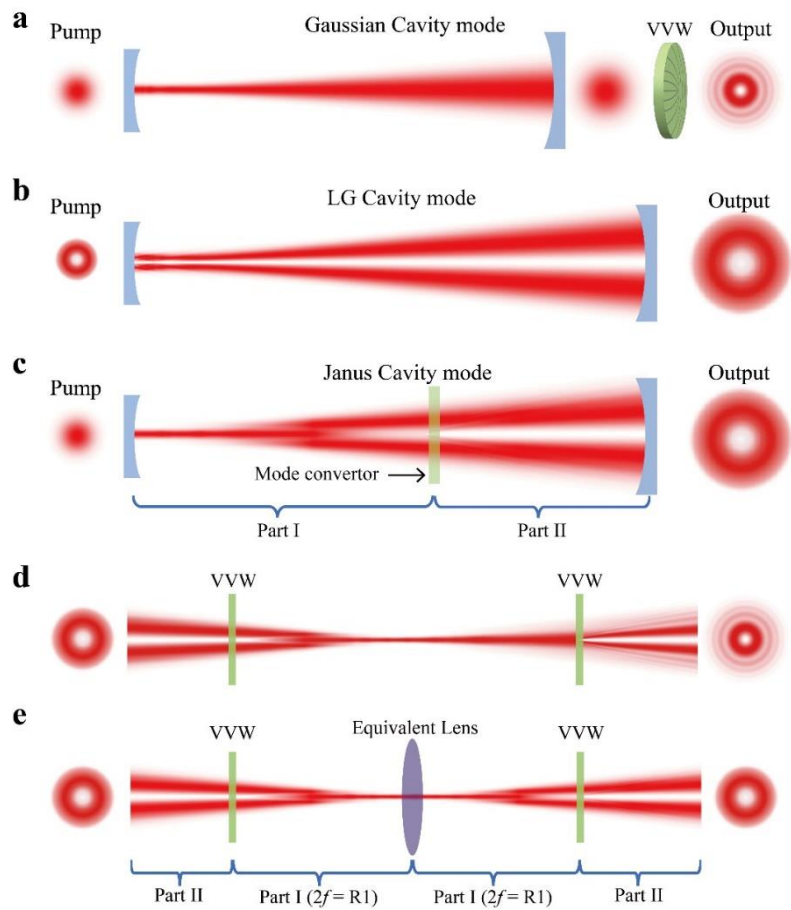

**Fig. 1 | Different cavity modes in OPO and Janus OPO designs. a.** A Gaussian-pumped OPO oscillating in a fundamental Gaussian mode. **b.** An LG-pumped OPO with an LG cavity mode and

an LG output mode. **c.** A specially-designed Janus OPO that is pumped by a Gaussian mode but outputs an LG mode. **d.** Evolution of a hollow-Gaussian beam. An LG mode passing through a VVW produces a hollow-Gaussian beam, which evolves into a Gaussian-like mode after a certain propagation. However, the hollow-Gaussian beam cannot recover itself without the equivalent lens as in **e** and neither can the LG mode. **e.** A one-round-trip mode conversion inside a Janus OPO. The input coupler with a radius curvature of R1 can be seen as an equivalent lens with a focusing length of $2f = R1$. Therefore, the light field after the VVW, which is set at the curvature center, will recover itself at the same position after reflected by the input coupler.

**Experimental setup of the Janus OPO**

Figure 2 shows the experimental setup of Janus OPO for generation of an LG-mode signal beam. Its output wavelength is designed to be tunable within the optical communication band. Two concave mirrors form the input and output couplers, which are coated for high reflectivity at the signal wavelength. A PPLN crystal serves as the nonlinear medium, which has multiple channels to extend the QPM bandwidth (see Supplementary Section 1 for details). The pump beam is generated by a 1064 nm pulsed nanosecond laser. Besides the Janus cavity mode as discussed above, the polarization of the field in the cavity is also precisely controlled to facilitate the parametric down conversion and mode conversion. In the PPLN crystal, both the pump and signal waves polarize vertically to utilize the biggest nonlinear coefficient $d_{33}$ of the PPLN crystal for high conversion efficiency. By changing temperature and selecting channel of the PPLN crystal, the output wavelength of signal wave can span from 1480 nm to 1650 nm (see Supplementary Section 1 for details). The next mode conversion sub-system includes a Faraday rotator (FR), a quarter-wave plate (QWP) and a VVW (the system has a work bandwidth of 1550 nm ± 50 nm). The VVW has a distinct $q$ factor with its value of being positive multiple of 1/2. In the forward propagation direction, a TC of $l = 2q$ is loaded onto the signal wave, which will be cancelled when it passes through the VVW again on its way back. However, due to the mismatched intensity distributions, it is impossible to directly transfer a Gaussian mode to an ideal LG mode through a VVW alone. In order to generate a high-purity LG mode at the output, the cavity mode in Part I should be re-configured to form a superposition of multiple spatial modes, which is automatically achieved by Janus cavity design (see Supplementary Section 2 for details). Here, we use a concave input coupler and set the VVW at its curvature center, which composes a symmetric imaging system to satisfy the condition of multi-mode reversibility in Part I of the Janus OPO (Fig. 1e). In principle, an ideal LG mode propagates in Part II of the cavity. Such stable Janus cavity mode is confirmed by a numerical calculation based on the Fox-Li simulating process (Fig. 1c) (See Method and Supplementary Section 2 for the details). It should be noted that the mode-conversion process requires a circularly-polarized signal wave on the VVW. To fulfill the polarization reversibility in the cavity, we add a FR and a QWP to accomplish the polarization control (see Supplementary Section 3 for details).

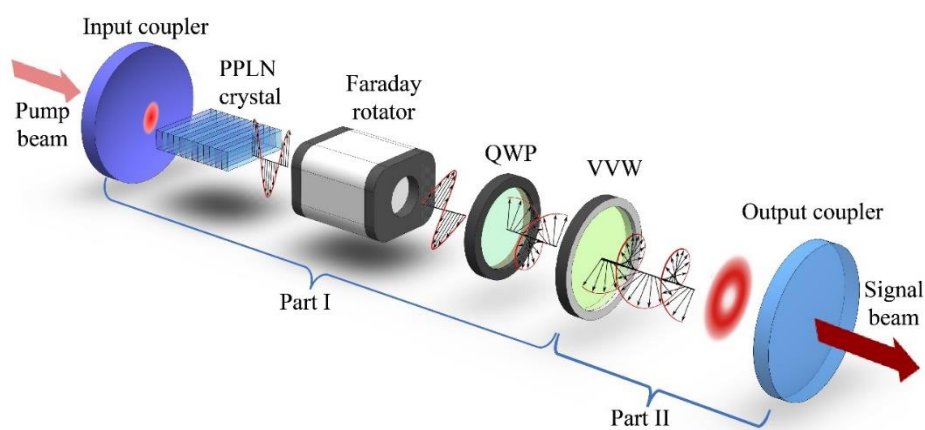

**Fig. 2 |Experimental setup**. The PPLN crystal, as the nonlinear medium, transforms one pump photon into a signal photon and an idle photon through the QPM parametric down-conversion process. The input/output couplers are coated for high reflectivity at the signal wavelength. The FR, QWP, and VVW form a mode conversion setup inside the cavity. The QWP alters the vertical polarization of the signal beam to circular polarization so that the spin-orbital angular momentum conversion could happen on the VVW to achieve the desired Gaussian-to-LG mode conversion. The output LG mode can be changed by rotating the QWP or replacing the VVW. FR is used to keep the signal wave to be vertically polarized inside the PPLN crystal.

**Performance of the Janus OPO**

First, we demonstrate highly-efficient generation of high-purity LG(1, 0) and LG(-1, 0) modes. A VVW with $q = 1/2$ is used to introduce a TC of $l = \pm 1$. The sign is controlled by the orientation of QWP. Under QPM configuration, the vertically-polarized pump beam produces a vertically-polarized signal beam in the PPLN crystal. After passing through the FR, the signal beam has a 45° linear polarization. When the fast axis of the QWP orients vertically (or horizontally), the signal polarization is further changed to a left- (or right-) circularly-polarized one, resulting in $l = 1$ (or -1) after the VVW (Fig. 2) (see Supplementary Section 3 for details).

Figure 3a shows the dependence of output powers of LG(1, 0) and LG(-1, 0) modes on the pump power. The output wavelength is set at 1550 nm. Both modes have a similar OPO threshold of ~1.2 W and a considerable sloping efficiency of 21.3%. For a pump power of 4.3 W, the output power reaches 660 mW and 648 mW for the LG(1, 0) and LG(-1, 0) modes, respectively. The measured conversion efficiencies of 15.3% and 15.1% are comparable to a typical Gaussian-mode OPO system. The intensity patterns of the output LG(1, 0) and LG(-1, 0) modes (Fig. 3b) exhibit high-quality donut intensity distribution without observable side lobes. Clearly, the undesired higher-order LG modes ($p > 0$) are significantly suppressed by the Janus OPO cavity. Further modal analysis in Fig. 3c shows that the generated LG(1, 0) and LG(-1, 0) modes have mode purities of 96.6% and 97.0%, respectively (see Supplementary Section 4 for modal analysis process). The mode purity is greatly enhanced in comparison to the typical value of ~80% using a VVW outside the cavity[40].

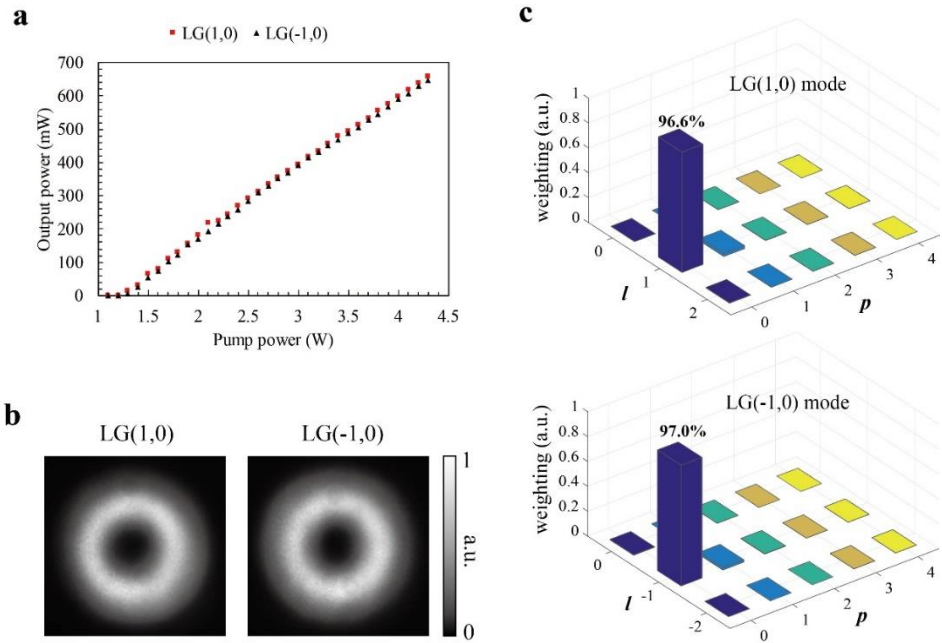

**Fig. 3 | Controllable generation of high-purity LG(1, 0) and LG(-1, 0) modes at the wavelength of 1550 nm. a.** Dependences of the output powers of the signal LG(1, 0) and LG(-1, 0) modes on the pump power show OPO thresholds at 1.2 W. The sloping efficiency is ~21.3%. **b.** The intensity

patterns of LG(1, 0) and LG(-1, 0) output modes exhibit well-defined donut-shaped intensity distributions. **c.** Modal analyses of the LG(1, 0) and LG(-1, 0) modes in **b** demonstrate high mode purities of 96.6% and 97.0%, respectively.

In our expeirment, the output wavelength of the Janus OPO can be tuned by changing QPM channel and temperature of the PPLN crystal. The Janus OPO shows an excellent performance within the designed wavelengths ranging from 1500 nm to 1600 nm. As shown in Fig. 4a, the conversion efficiency of the signal LG mode surpasses 10% in the most of the working wavelengths. Figure 4b compares the power dependence of the output LG(1, 0) mode on the pump power at 1525 nm, 1550 nm, 1575 nm, and 1600 nm, respectively. The differences in threshold and conversion efficiency for different wavelengths can be attributed to the fact that the intra-cavity optical components are not uniformly optimized at all the wavelengths. Figure 4c depicts the modal analysis results of the output LG(1, 0) mode at the wavelengthes of 1525 nm and 1575 nm, which show high mode purities of 97.1% and 95.9%, respectively. The bandwidth of this Janus OPO can be further extended by using ultra-wide-band opitcal components as intra-cavity elements.

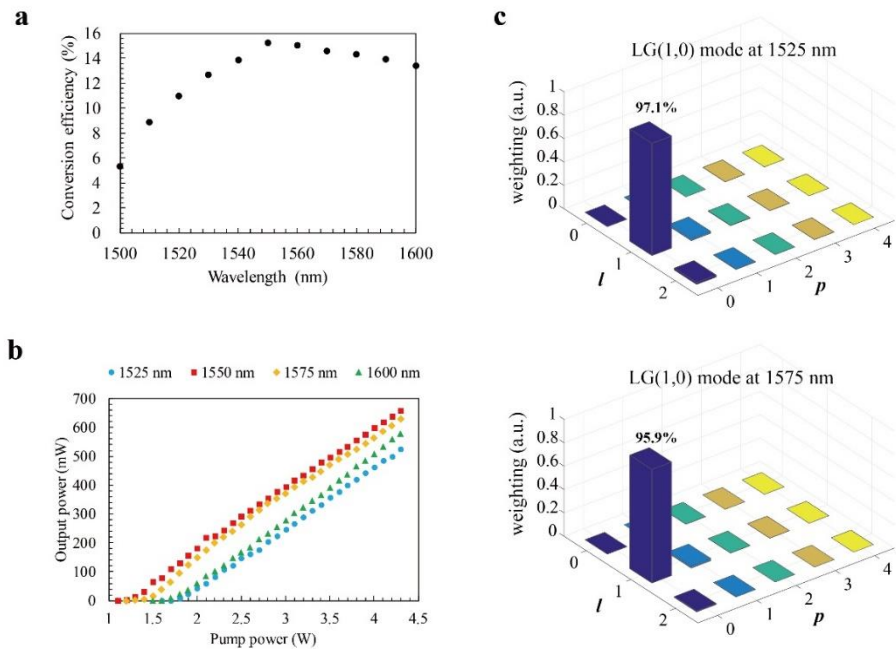

**Fig. 4 | Wavelength tunable high-purity LG(1, 0) mode. a.** Dependence of the conversion efficiency of the signal LG mode on the wavelength ranging from 1500 nm to 1600 nm, showing

high conversion efficiency at the designed bandwidth. **b.** Dependences of output powers of the signal waves on the pump power at 1525 nm, 1550 nm, 1575 nm, 1600 nm, respectively, showing high-quality OPO output performances. **c.** Modal analysis for the output LG(1, 0) mode at the wavelengths of 1525 nm and 1575 nm, showing the mode purity up to 97.1%.

Our specially-designed Janus OPO is also suitable for generating high-order LG modes. As a demonstration, VVWs of $q = 1$ and $q = 2$ are used to generate LG($\pm 2$, 0) and LG($\pm 4$, 0) modes. Since LG($\pm l$, 0) modes experience similar OPO process, we only show the results of LG(2, 0) and LG(4, 0) modes. Figure 5a shows the conversion efficiencies at the wavelength range between 1500 nm and 1600 nm. The maximal conversion efficiencies for the LG(2, 0) and LG(4, 0) modes at 1550 nm are 15.8% and 15.6%, respectively. High conversion efficiencies beyond 10% is achieved over an 80 nm wavelength bandwidth. The mode purities of 95.2% for the output LG(2, 0) mode and 93.7% for the LG(4, 0) mode are much superior as compared to 60% and 50% values using VVWs outside the cavity[40]. Note that the performance of such Janus OPO does not degrade much for outputs of high-order LG modes. The design of Janus cavity fully takes advantages of two important facts, i.e., its Gaussian face for high-efficiency frequency conversion and LG face for high-purity LG output. Our simulations also show smooth Janus cavity modes for different LG mode orders (see Supplementary Section 3 for details).

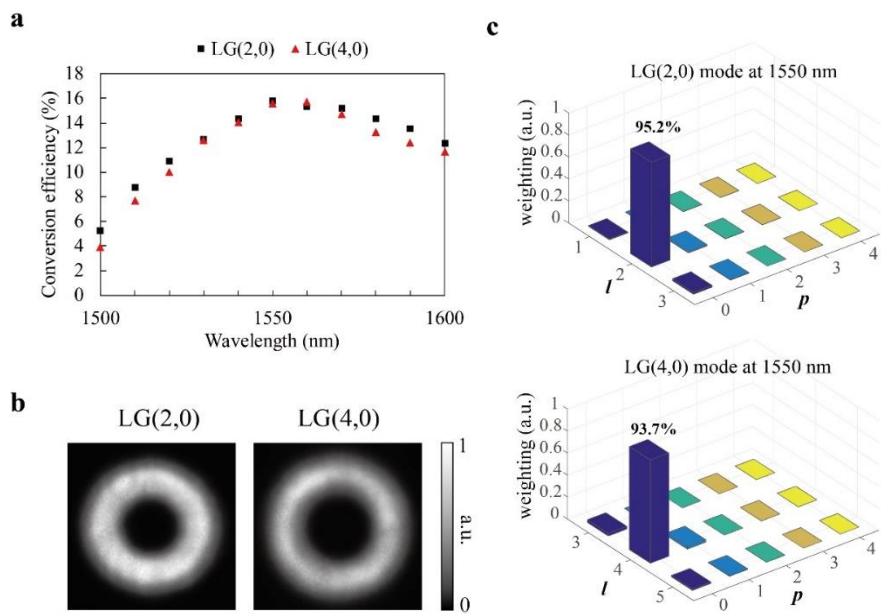

**Fig.5 | High-order LG(*l*, 0) modes with TCs of *l* = 2 and *l* = 4. a.** Dependences of conversion efficiencies on output wavelengths for LG(2,0) and LG(4,0) modes, respectively, at a pump power of 4.2 W. Over 10% conversion efficiency covers the wavelength range from 1520 nm to 1600 nm. **b.** Intensity patterns of the LG(2, 0) and LG(4, 0) modes at the wavelength of 1550 nm. **c.** Modal analyses of the output LG(2, 0) and LG(4, 0) modes in **b**, giving mode purities of 95.2% and 93.7% respectively.

**Conclusion**

We have experimentally demonstrated a Janus OPO system for generating highly-efficient, highly-pure, broadly-tunable, and TC-controllable LG modes. Such Janus OPO distinguishes itself by possessing the two-face cavity mode, which makes use of the distinct advantages of both the Gaussian and LG cavity modes. The front (input) face has a Gaussian profile to achieve the high-efficiency nonlinear frequency conversion while its back (output) face is a donut-shaped LG profile that guarantees the direct output of a desired high-purity LG mode from the cavity. The key to realize such a Janus OPO is the introduction of an imaging system to facilitate the perfect intra-cavity mode conversion. In this work, the Janus OPO is designed for the Gaussian-to-LG mode conversion of the signal light, which can be easily adjusted to output an LG mode at the idler wavelength. The conversion efficiency of Janus OPO could be further enhanced by use of a double-pass pump configuration[49]. In addition, by selecting proper optical components, our experimental configuration can be readily extended to visible and UV wavelength bands, as well as to generate tunable vector beams and multi-dimensional quantum entangled sources. The excellent performance features of the LG modes from our Janus OPO (e.g. wavelength tunable between 1.5 μm and 1.6 μm, conversion efficiency > 15%, and mode purity > 97%) can meet the critical requirements of high-level applications such as high-capacity optical communications, high-precision sensing and measurements, and super-resolution imaging.

**Methods**

**Experimental setup of the Janus OPO.** As shown in Fig. 2a, a PPLN crystal with dimensions of $25(x) \times 12.3(y) \times 1(z)$ mm$^3$ serves as the nonlinear medium. Both of its end faces have transmittance >99% at 1380-1800 nm wavelength range. It is mounted inside an oven with the temperature tunability up to 150 °C. The input coupler (with a radius of curvature of -75 mm) is coated with a high transmittance (>99%) at 1064 nm and a high reflectivity (>99%) at 1450-1650 nm, while the output coupler (with a radius of curvature of -125 mm) is coated with a transmittance of 30% at 1450-1650 nm. The cavity length is 140 mm, satisfying the stability condition of a resonator. A pump beam (wavelength 1064 nm, repetition rate 22 kHz, pulse width 45 ns) is generated by a nanosecond pulsed fiber laser (YDFLP-M7-3-PM, JPT Co.). It is focused by a lens into a 200 μm-in-diameter spot inside the crystal. The PPLN crystal has ten channels. In the experiment, we use four channels with periods of 31.02 um, 30.49 um, 29.98 um, and 29.52 um, respectively. Under the pumping wavelength of 1064 nm, the output signal wavelength can be tuned from 1480 nm to 1650 nm in the temperature range from 25 °C to 138 °C (see Supplementary Section 1 for details). A Faraday rotator (FR), a quarter-wave plate (QWP) and a vector vortex waveplate (VVW) are inserted into the cavity to achieve the reversible mode conversion inside the cavity. All their working wavelength bandwidths are from 1500 nm to 1600 nm. The VVW is placed at a distance of 90 mm away from the input coupler, where is the curvature center of the input coupler considering the effective length due to the high refractive index of the PPLN crystal. VVWs of $q = 0.5, 1, 2$ have been used to generate LG($l$, 0) modes with different TCs. The output intensity patterns are recorded by a laser beam profiler (LBP, Newport Corp.)

**Cavity mode simulations.** The numerical simulations have been carried out based on Fox-Li method. A one-round-trip transition of the cavity mode can be described as following. A parametric wave starting from the input coupler travels a distance of $L_A$ and passes through the VVW with a TC of $l$ (or $-l$). After travelling a distance of $L_B$, the parametric wave is reflected by the output coupler and propagates backward. The TC is cancelled when the parametric wave passes through the VVW along the opposite direction. Finally, the parametric wave reaches the input mirror to finish its one-round-trip transition. The parametric wave repeats the cycle until a stable cavity mode is formed. The iterative procedure described above is calculated step by step by using Matlab programming.

**Acknowledgements**

This work was supported by the National Key R&D Program of China (2017YFA0303703 and

**Additional information**

The authors declare no competing interests.

# Supplementary Information

1. **The dependence of output signal wavelength on PPLN channel and temperature.**

A traditional optical parametric oscillator (OPO) using a PPLN crystal converts a pump wave into a signal wave and an idle wave. The three waves satisfy the energy conservation, i.e., $\omega_p = \omega_s + \omega_i$, where $\omega_p$, $\omega_s$, and $\omega_i$ refer to the frequencies of the pump, signal, and idle waves, respectively. By changing the temperature and channel of the PPLN crystal, $\omega_s$ and $\omega_i$ can be tunable due to the momentum conservation condition $\vec{k}_p - \vec{k}_s - \vec{k}_i - \vec{G} = 0$, where $\vec{k}_p$, $\vec{k}_s$, and $\vec{k}_i$ refer to the wave vectors of the pump, signal, and idle waves, respectively [S1,S2]. $\vec{G}$ is the first-order reciprocal vector provided by the PPLN crystal. In our experimental setup, the PPLN crystal has ten channels. We use four channels with periods of 31.02 μm, 30.49 μm, 29.98 μm, and 29.52 μm, respectively, as shown in Fig. S1. Under the pump wavelength of 1064 nm, the output signal wavelengths can be tuned from 1480 nm to 1650 nm in the temperature ranging from 25°C to 138°C (Fig. S2).

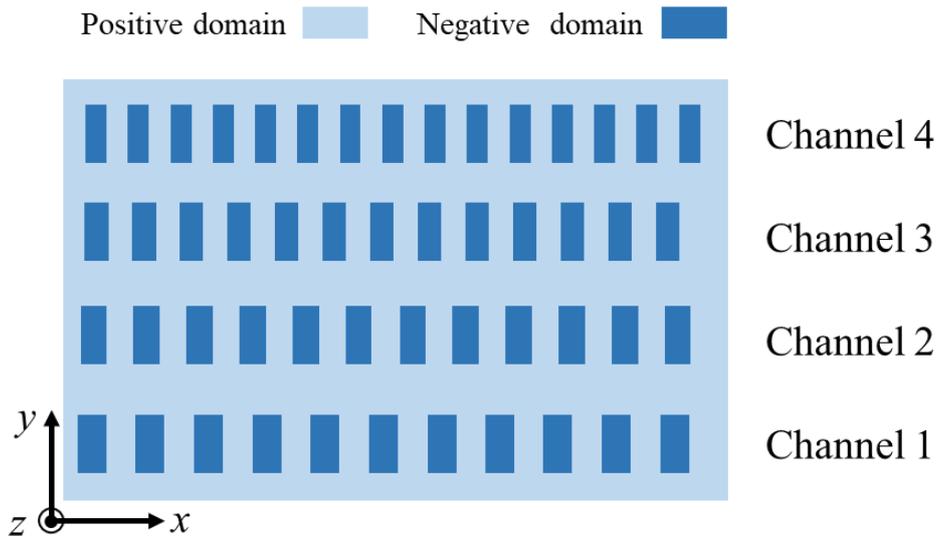

**Fig. S1** | Schematic image of four channels in the PPLN crystal. The periods are 31.02 μm, 30.49 μm, 29.98 μm, and 29.52 μm, corresponding to Channel 1, 2, 3, and 4, respectively.

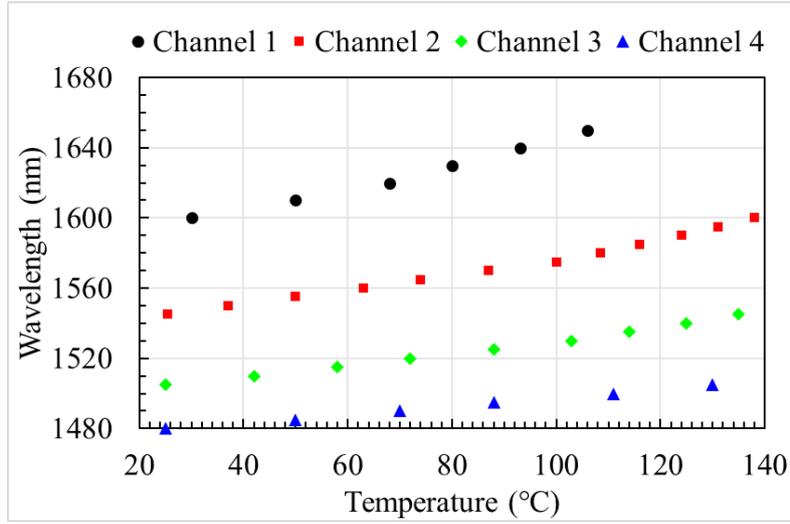

**Fig. S2** | The dependence of output signal wavelength on PPLN channel and temperature.

## 2. Janus cavity mode simulation

Based on the Fox-Li method, we simulate the Janus cavity mode with VVWs of $q$ = 0.5, 1, and 2, which can generate LG(1,0), LG(2,0), and LG(4,0) modes, respectively. The one-round-trip transitions of the simulated stable Janus cavity mode are shown in Fig. S3. The symmetric imaging system facilitates the cavity mode to smoothly evolve from a Gaussian profile to an LG profile, and vice versa, without breaking the cavity mode reversibility. The simulated output modes show high-quality donut intensity distribution without observable side lobes.

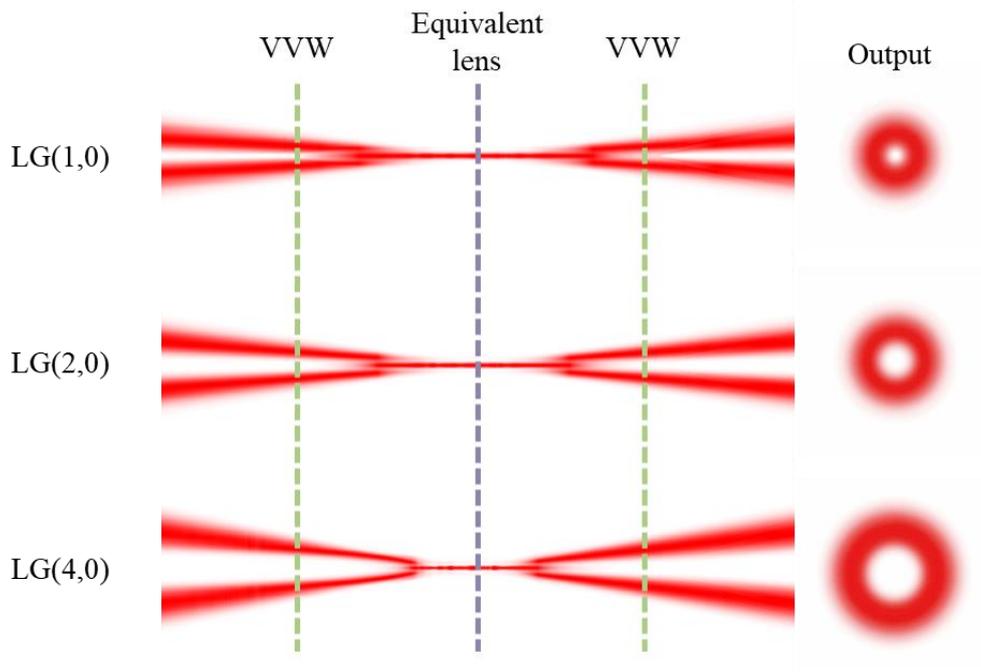

**Fig. S3** | One-round-trip mode conversions of the stable Janus cavity modes for generating LG(1,0), LG(2,0), and LG(4,0) modes, respectively.

3. **Reversible polarization conversion**

A VVW can be seen as a spatially variant half-wave plate, whose optical axis rotates continuously around a singularity point. Its transmissivity is up to 95% covering wavelengths from 1500 nm to 1600 nm. The orientation of its fast axis can be expressed as:

$$\theta(\varphi)=q\varphi+\varphi_0, \tag{S1}$$

where $\varphi$ is the variant azimuthal angle, $\varphi_0$ is the orientation of the fast axis at $\varphi=0$ and $q$ is a constant with its value of being positive multiple of 1/2. Its Jones Matrix can be written as:

$$J_q(\theta)=\begin{bmatrix} \cos 2\theta & \sin 2\theta \\ \sin 2\theta & -\cos 2\theta \end{bmatrix}. \tag{S2}$$

When a circularly polarized beam passes through the VVW, the following transformation happens [S3],

$$\begin{cases} J_q(\theta) \times |l_0, \ L\rangle = |l_0+m, \ R\rangle \\ J_q(\theta) \times |l_0, \ R\rangle = |l_0-m, \ L\rangle \end{cases}, \tag{S3}$$

where $L$ and $R$ refer to the left-circularly-polarized (LCP) and right-circularly-polarized (RCP) states, respectively. From Eq. (S3), we can see that a LCP (or RCP) state with TC of $l_0$ passing through the VVW will become a RCP (or LCP) state with TC of $l_0+2q$ (or $l_0-2q$). When a $l_0 = 0$ input state is used, the TC will be controlled by the $2q$ value of the VVW and the handedness of the incident circularly-polarized beam.

In the experiment, a broadband quarter-wave plate (QWP) is used to convert the generated linearly-polarized signal wave into a circularly-polarized one. The Jones matrix of QWP can be expressed as

$$G_{\pi/2}(\beta) = \frac{\sqrt{2}}{2}\begin{bmatrix} 1-i\cos 2(\beta-\beta_0) & -i\sin 2(\beta-\beta_0) \\ -i\sin 2(\beta-\beta_0) & 1+i\cos 2(\beta-\beta_0) \end{bmatrix}. \quad (S4)$$

where $\beta$ and $\beta_0$ is the orientation angle of the fast axis of QWP and the polarization angle of the signal wave, respectively. Consider that a light beam propagates forward through QWP and VVW, is reflected by the output coupler, and then propagates backward through VVW and QWP. The process can be expressed by,

$$\begin{aligned}J_{total}(\theta,\ \beta) &= G_{\pi/2}(-\beta)J_q(-\theta)J_m J_q(\theta)G_{\pi/2}(\beta) \\ &= -2i\begin{bmatrix} \cos 2(\beta-\beta_0) & \sin 2(\beta-\beta_0) \\ -\sin 2(\beta-\beta_0) & \cos 2(\beta-\beta_0) \end{bmatrix}.\end{aligned} \quad (S5)$$

$J_m$ is the Jones Matrix describing the mirror reflection, which is given by

$$J_m = \begin{bmatrix} 1 & 0 \\ 0 & -1 \end{bmatrix}. \quad (S6)$$

Note that the right-hand coordinate system requires the azimuthal angle to change its sign when the light wave propagates backward.

It requires $\beta - \beta_0 = \pm 45°$ for transforming a linearly-polarized state into a circularly-polarized state. So one can obtain from Eq. (S5) that the polarization is rotated by $\pm 90°$ after the light passes through QWP and VVW forward and backward. We use a Faraday rotator (FR) to compensate the $\pm 90°$ polarization rotation so that the total polarization conversion process is reversible as shown in Fig. S4.

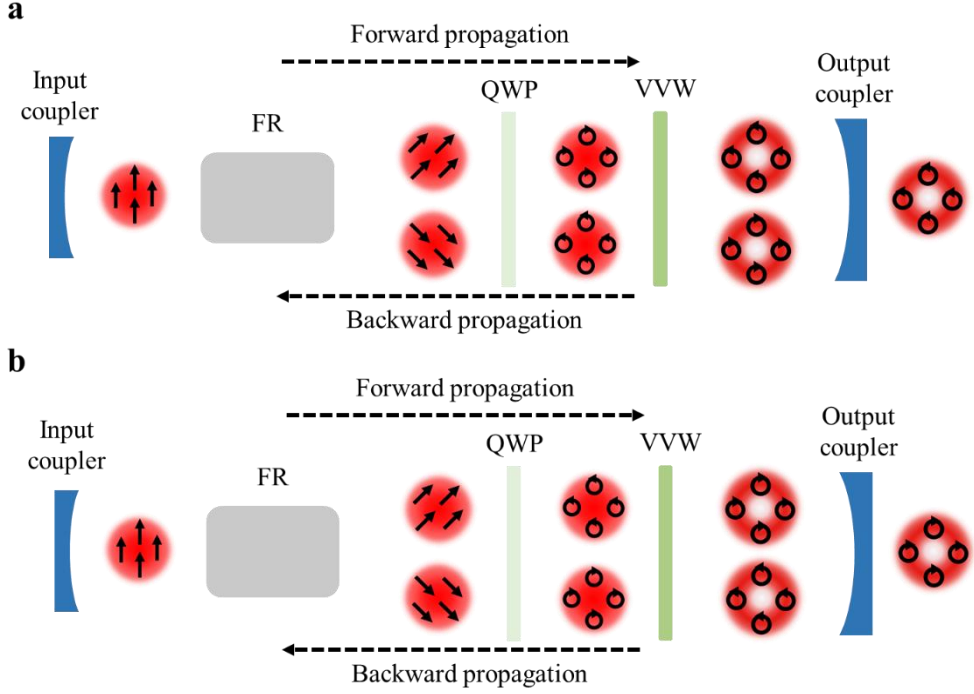

**Fig. S4** | Polarization and mode conversions inside the Janus OPO cavity for outputs of (**a**) right-circularly-polarized LG(2q,0) mode and (**b**) left-circularly-polarized LG(-2q,0) mode. The value of q is positive multiple of 1/2. The solid arrows indicate the polarization while the dash arrows show the propagation directions.

4. **Modal analysis process.**

Modal decomposition using digital holograms is a common way to characterize the spatial beams [S4]. Because all the $LG_l^p$ modes construct a complete and orthonormal basis, the spatial mode $U(x, y)$ can be expanded into the coherent superposition of $LG_l^p$ modes:

$$U(x,y) = \sum_{l=-\infty}^{\infty} \sum_{p=0}^{\infty} c_l^p LG_l^p(x,y). \quad (S7)$$

Here, $c_l^p$ is the weighting coefficient, which can be calculated from

$$c_l^p = \langle LG_l^p(x,y) | U(x,y) \rangle = \iint U(x,y) LG_l^p(x,y)^* dxdy. \quad (S8)$$

$LG_l^p(x, y)^*$ is the complex conjugate of $LG_l^p(x, y)$. To simplify the calculation, we apply a Fourier transform on $U(x, y) \cdot LG_l^p(x, y)^*$,

$$U_k(k_x, k_y) = \iint U(x,y) LG_l^p(x,y)^* \exp[-i(k_x x + k_y y)] dxdy, \quad (S9)$$

where the $k_x$ and $k_y$ are the wave vectors. Then, the on-axis optical field in the Fourier plane is

given by

$$U_k(0,0) = \iint U(x,y) LG_l^p(x,y)^* \, dxdy. \tag{S10}$$

Therefore, by measuring the intensity of $U_k(0,0)$, the power weighting of the corresponding $LG_l^p$ component can be determined by

$$\left|c_l^p\right|^2 = \left|U_k(0,0)\right|^2. \tag{S11}$$

Equation (S11) provides a practical way to determine the mode purity of the spatial mode.

In the experiment, we use a reflective phase-only spatial light modulator (SLM, GAEA-2 - TELCO, HOLOEYE Corporation) to perform the modal analysis. We use the type-3 method of complex-amplitude modulation reported by Arrizon *et al.* to program the phase-only computer-generated holograms (CGH) [S5]. A spatial carrier frequency is added into the CGH so that the 1st order diffraction beam (which reconstructs the conjugate of the tested $LG_l^p$ component) is separated from the undesired modes in the Fourier plane.

Figure S5 shows the schematic setup for the modal decomposition. Because the SLM is only valid for linearly-polarized light, we turn the output circularly-polarized LG mode into a linearly-polarized one by using a QWP. Then it passes through a 1:1 beam splitter (BS) and is incident on the SLM. The light field reflected at SLM and BS in sequence carries the information of $U(x,y) LG_l^p(x,y)^*$, which is focused by a 75 mm lens to perform the Fourier transform. At the Fourier plane, the first-diffraction-order beam reconstructs the field $U_k(k_x, k_y)$, which is picked out through an iris and is imaged on the LBP by a 50X objective. By recording the central intensity of the image, the power weighting is achieved. In the experiment, we measure the modal distribution of the generated $LG_1^0$ beam by loading a group of $LG_l^{p*}$ modes with $p$ varying from 0 to 4 and $l$ varying from 0 to 2, onto the SLM.

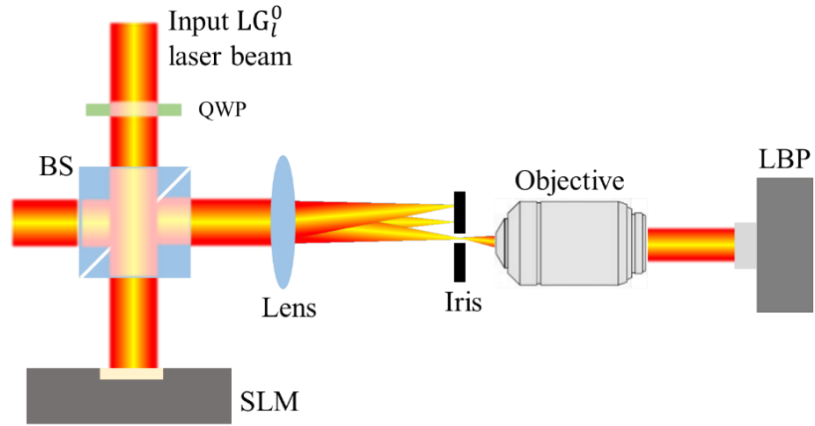

**Fig. S5 |** The schematic setup for modal decomposition.